\documentclass[aps,prd,twocolumn,floatfix,superscriptaddress,balancelastpage,%
nofootinbib,amsmath]{revtex4}

\usepackage{graphicx}
\usepackage{epsfig}
\usepackage{bm}

\newcommand{\lwig}{\mbox{\;\raisebox{.3ex}
    {$<$}$\!\!\!\!\!$\raisebox{-.9ex}{$\sim$}\;}}
\newcommand{\gwig}{\mbox{\;\raisebox{.3ex}
    {$>$}$\!\!\!\!\!$\raisebox{-.9ex}{$\sim$}}\;}

\begin{document}

\hspace*{145mm} {\large \tt DESY~05-053}

\title{$Z$-bursts from the Virgo cluster}

\author{Andreas Ringwald}
\email{andreas.ringwald@desy.de}
\affiliation{Deutsches Elektronen-Synchrotron DESY, Hamburg, Germany}

\author{Thomas J.~Weiler}
\email{t.weiler@vanderbilt.edu}
\affiliation{Department of Physics and Astronomy, Vanderbilt University, 
Tennessee 37235, USA}

\author{Yvonne Y.~Y.~Wong}
\email{yvonne.wong@desy.de}
\affiliation{Deutsches Elektronen-Synchrotron DESY, Hamburg, Germany}

\begin{abstract}
Resonant annihilation of ultra-high energy cosmic neutrinos (UHEC$\nu$)
on the cosmic neutrino background (C$\nu$B) into $Z$ bosons---the
$Z$-burst mechanism---and its associated absorption and emission phenomenology
 provide a unique, albeit indirect, probe of the C$\nu$B  in its present state.
   In this paper, we examine the implications of gravitational
clustering of the C$\nu$B in nearby galaxy clusters for the $Z$-burst
phenomenology.  In particular, we study the emission features of the $Z$-decay
products originating from the Virgo cluster, and the potential of future cosmic
ray experiments to observe clustering-enhanced $Z$-burst rates.  We find that a
detector with an exposure equivalent to three years of observations at the
Extreme Universe Space Observatory (EUSO) will very likely measure these
enhanced rates together with the associated UHEC$\nu$ flux, provided that the
latter saturates current observational limits and  the neutrino masses are
quasi-degenerate, $m_{\nu_i} \gwig 0.1 \ {\rm eV}$.  In the case of
UHEC$\nu$ fluxes below
the electromagnetic cascade limit, or a hierarchical neutrino mass spectrum,
an experimental sensitivity exceeding that of
EUSO by at least two orders of magnitude is required to detect the clustering
enhancements with any certainty.
\end{abstract}
\maketitle

\section{Introduction}

The existence today of a $1.95$ K cosmic neutrino background (C$\nu$B)---an
exact analogue of the $2.73$ K cosmic microwave background (CMB)---is a
fundamental prediction of the standard big bang theory.  Permeating the
universe at an average number density of $\bar{n}_{\nu_i} =
\bar{n}_{\bar{\nu}_i} \simeq 56 \ {\rm cm}^{-3}$ per flavour, these relics
of the big bang trace their origin to the freeze-out of the weak
interaction when the universe was a mere one
second old ($T \sim 1 \ {\rm MeV}$), predating even the CMB photons by
thirteen orders of magnitude in time. Yet, for the same reason that
they decoupled so early, the C$\nu$B neutrinos have so far escaped direct
detection in a controlled laboratory setting. To date, cosmological
measurements such as the CMB anisotropies and the large-scale matter power
spectrum, and, independently, the observed light elemental abundances,
provide the best probe of the C$\nu$B's presence in the {\it early stages} of
cosmological evolution (e.g., \cite{bib:hannestadreview}).
It is therefore our view that alternative avenues, however indirect, that could
potentially afford us a glimpse of the C$\nu$B
{\it as it is today}  should be thoroughly explored.

One such avenue is the $Z$-dip/burst mechanism and its associated phenomenology
\cite{Weiler:1982qy}, which proposes to exploit the resonant annihilation of
hypothetical ultra-high energy cosmic neutrino (UHEC$\nu$) beams on the C$\nu$B
as a target. Supposing such beams exist, the  annihilation process
$\nu_{\rm UHEC\nu} + \bar{\nu}_{\rm C \nu B} \to Z \to {\rm hadrons}$
proceeds at the resonance energy $E^{\rm res}_{\nu_i}$ with a cross section
enhanced by several orders of magnitude compared to non-resonant scattering, and
\begin{equation}
E^{\rm res}_{\nu_i} = \frac{m_Z^2}{2 m_{\nu_i}} = 4.2 \times 10^{21}
\left(\frac{\rm eV}{m_{\nu_i}} \right) \ {\rm eV}
\end{equation}
is a function of the neutrino and the $Z$  masses, $m_{\nu_i}$ and $m_Z$, alone.
The annihilation can be detected as absorption dips in the
incident UHEC$\nu$
flux at $E \sim E^{\rm res}_{\nu_i}$ (``$Z$-dips'')
\cite{Weiler:1982qy,Roulet:1993pz} and/or as emission features in the $Z$-decay
products (nucleons and photons) \cite{Fargion:1999ft}.
Indeed, in the latter case, the happy coincidence between $E_{\nu_i}^{\rm res}
\sim 10^{21} \ {\rm eV}$ (for
$m_{\nu_i} \sim 1 \ {\rm eV}$)
and the energies of the most energetic cosmic rays observed by AGASA
\cite{bib:agasa}, Fly's Eye \cite{bib:flyseye}, HiRes \cite{bib:hires}, Yakutsk
\cite{Yakutsk}, and Haverah Park \cite{HP}
has long led to a possible identification of ultra-high energy cosmic rays
(UHECR) above the Greisen--Zatsepin--Kuzmin (GZK) cut-off energy $E_{\rm GZK}
\sim 4 \times 10^{19} \ {\rm eV}$ \cite{Greisen:1966jv} with $Z$-burst nucleons
and photons \cite{Fargion:1999ft,Kalashev:2001sh}.

\begin{figure}[t]
\epsfxsize=8.5cm
\epsfbox{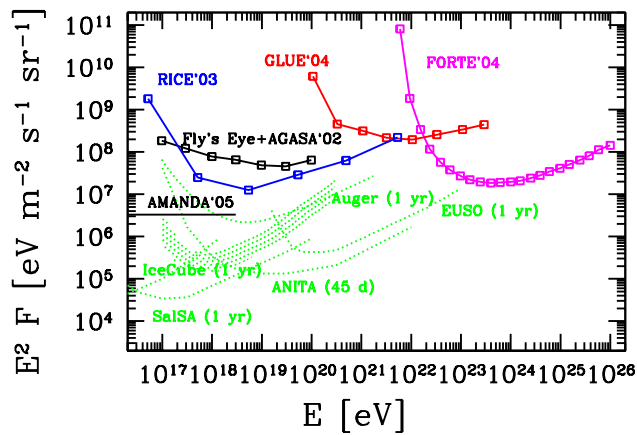}
\caption{Upper limits on the diffuse neutrino flux per flavour
$F_{\nu_{\alpha}}+F_{\bar{\nu}_{\alpha}},\, \alpha=e,\mu,\tau$,
from RICE \cite{bib:rice}, GLUE \cite{bib:glue},
Fly's Eye+AGASA~\cite{bib:flyseye+agasa}, FORTE \cite{bib:forte}
and AMANDA (assuming an
$E^{-2}$ differential spectrum) \cite{Ackermann:2005sb}, assuming
full flavour mixing en route to Earth.
Also shown are the projected sensitivities of Auger in
$\nu_e, \ \nu_{\mu}$ modes and in $\nu_{\tau}$ mode (bottom swath)
\cite{bib:auger}, ANITA \cite{bib:anita}, EUSO \cite{bib:euso}, IceCube
\cite{bib:icecube}, and SalSA \cite{bib:salsa}, i.e., one event per energy decade
for the indicated duration.  A prototype of the full ANITA experiment,
ANITA-lite \cite{bib:anita-lite}, should constrain the $10^{19} \div 10^{22} \
{\rm eV}$ region in the very near future.\label{fig:current_future}}
\end{figure}

\begin{figure}[t]
\epsfxsize=8.5cm
\epsfbox{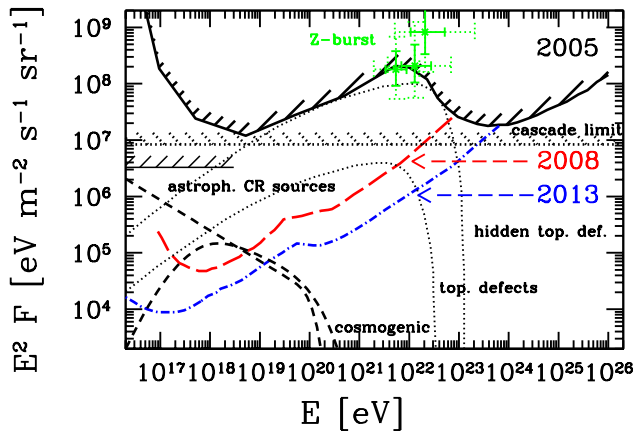}
\caption{Prospects for ultra-high energy neutrino detection in the next decade
(fluxes required  to produce one event per energy decade).
For the year 2008 (long dashed/red line),
we assume three years of Auger data and one 15-day ANITA flight.
For 2013 (dash-dot/blue line), we assume 8/3/3/4 yr
Auger/EUSO/IceCube/SalSA, and three ANITA flights.  These curves are indicative 
only.  The sensitivity will improve if we consider also the Westerbork radio
observatory \cite{Westerbork}, if further projects such as Auger North and OWL
\cite{bib:owl} are realised, or if the EUSO or ANITA flight times
are extended.  Also shown are a wide sample of UHEC$\nu$ flux predictions,
the current observational upper bound (solid shade)
from Figure \ref{fig:current_future}, and the electromagnetic cascade limit
from EGRET \cite{Sreekumar:1997un,Mannheim:1998wp}. The points labelled 
``$Z$-burst'' (green) denote the UHEC$\nu$ fluxes required to explain the
post-GZK events observed by AGASA in terms of $Z$-burst secondaries 
\cite{Fodor:2002hy}.\label{fig:road}}
\end{figure}

Clearly, the success of the $Z$-burst mechanism as a means to detect the C$\nu$B
and/or as an explanation of UHECR depends first and foremost on the UHEC$\nu$
fluxes (the beam);  Figures \ref{fig:current_future} and
\ref{fig:road}
summarise the current status of the search for such fluxes, the projected
sensitivities of ongoing and planned experiments,
and predictions from various theoretical models.  A second factor is the nature
of the C$\nu$B density distribution (the target).
To this end, it is important to note that an oscillation interpretation of the
atmospheric and solar neutrino data (e.g., \cite{bib:atmospheric})
implies that at least two of the mass eigenstates in the C$\nu$B are
nonrelativistic  today, i.e., $m_{\nu_i} \gg T_\nu \sim (4/11)^{1/3}\,T_\gamma
\sim 2 \ {\rm K} \sim 10^{-4}\,{\rm eV}$, and at least one mass must exceed
$\sqrt{\Delta m^2_{\rm atm}}\sim 0.05\,{\rm eV}$. These nonrelativistic neutrinos
are subject to gravitational clustering on existing cold dark matter (CDM) and
baryonic structures, possibly causing the local C$\nu$B density to depart from
the cosmological average.

Standard large-scale structure theories
tell us that, in the currently favoured $\Lambda$CDM cosmology,
$\{\Omega_m,\Omega_{\Lambda},h\}=\{0.3,0.7,0.7\}$, fluctuations in the C$\nu$B
ought to track their CDM counterparts at scales above the neutrino free-streaming
length (e.g., \cite{bib:hannestadreview}).  The inferred local large-scale matter
distribution from peculiar velocity measurements \cite{bib:peculiar} therefore
precludes any increase in the relic neutrino content of the local GZK zone
($\sim 50 \ {\rm Mpc}$) by more than a factor of two due to
gravitational clustering alone \cite{Ringwald:2004np}.
However, clustering in local gravitational
potential wells
such as galaxies and galaxy clusters may still be sizeable on the sub-Mpc to Mpc 
scale.  Indeed, for neutrino masses satisfying experimental and cosmological
bounds,\footnote{
Limits on the neutrino mass derived from cosmological measurements, especially 
the matter power spectrum $P(k)$,
depend on the data sets used and the priors assumed.  For example, analyses 
including also the Lyman $\alpha$ data and information
on the dark matter--galaxy bias tend to produce bounds that are much tighter 
than those derived from the shape of $P(k)$ alone.
Furthermore, the neutrino mass appears in combination with several other 
uncertain cosmological parameters in the determination of $P(k)$, such as
a running scalar spectral index $n_s(k)$
and the effective number of thermalised fermionic degrees of freedom $N_{\nu}$.
These degeneracies can lead to considerable relaxation in
the bound on $m_{\nu_i}$.
In the present work, we take a
conservative upper limit of $\sum_i m_{\nu_i} < 1.8 \ {\rm eV} \ (2 \sigma)$ 
\cite{bib:sdss1} for three degenerate neutrinos,
derived from the SDSS  galaxy power spectrum  \cite{bib:sdss} and the WMAP data 
\cite{bib:wmap} assuming a constant $n_s$ and
$N_{\nu}=3$.
Laboratory bounds from tritium $\beta$-decay experiments, 
$m_{\nu} \lwig 2.2 \ {\rm eV}$  \cite{bib:mainz},
and from neutrinoless double beta decay, $m_{\nu}\lwig (0.66 \div 2.70 
)$~eV~\cite{bib:nu02beta}, are not yet
competitive.
See the reviews \cite{bib:hannestadreview,Elgaroy:2004rc}.} detailed 
calculations show that the C$\nu$B overdensities in the largest galaxy clusters 
($\sim 10^{15} \ M_{\odot}$) can be as much as a thousand within the central
$\sim 100 \ {\rm kpc}$ region \cite{Ringwald:2004np,bib:singh&ma}.  An immediate
consequence for the $Z$-burst scenario is a possible directional dependence in
its emission features, even if the UHEC$\nu$ sources are isotropically 
distributed;  the highest number of $Z$-burst events should originate from the
directions of nearby galaxy clusters.

The implications of gravitational neutrino clustering  for the $Z$-burst
emission spectra were first investigated quantitatively in reference
\cite{bib:singh&ma}.  In the present paper, we extend the said analysis in
several ways:
\begin{enumerate}
\item We present a more accurate determination of the C$\nu$B overdensities in
and around galaxies and galaxy clusters based on the calculations of reference
\cite{Ringwald:2004np}, which take into account nonlinear effects in the 
clustering process.  The linearised method adopted in  \cite{bib:singh&ma}
systematically underestimates the overdensities by a factor of 
several particularly in the virialised region of large galaxy clusters.

\item We compute both the primary nucleon and photon spectra.
A generic feature of the $Z$-burst scenario is a
 photon to nucleon ratio of almost 20:1 at the $Z$ production site.
This ratio is greatly reduced upon arrival at Earth because
of the much shorter attenuation length for photons
than for nucleons.
Still, unless the photons are
heavily attenuated by a strong universal radio background (URB) and/or by strong
extragalactic magnetic fields ($\gwig 10^{-10} \ {\rm G}$),
 we expect a predominance of photons in the observed primaries.

\item As in \cite{bib:singh&ma},
we assess the experimental prospects for observing enhanced $Z$-burst rates
from nearby galaxy clusters, focussing on our nearest neighbour: Virgo.
At an average distance of $\sim 15 \ {\rm Mpc}$ from the Milky Way and with a
mass close to $10^{15} M_{\odot}$ \cite{bib:virgomass},
the Virgo cluster is able to accumulate a sizeable excess of relic neutrinos 
while sitting close enough such that the $Z$-burst primaries suffer minimal 
energy loss during their propagation to Earth.  If clustering-enhanced 
$Z$-bursts were to be seen at all, the Virgo cluster would be the prime site.

\end{enumerate}

For rate estimates, we shall take the
Extreme Universe Space Observatory (EUSO) experiment \cite{bib:euso1}
as our fiducial detector.
The sensitivity of EUSO to UHECR and UHEC$\nu$ above $10^{20}$~eV
is about three orders of magnitude beyond what is available to date from AGASA 
and HiRes. As presently designed, EUSO will consist of a two-metre Fresnel lens
positioned on the International Space Station at a height
of 400 km.  The lens will focus near-UV fluorescence emitted by
radial de-excitation of $N_2$ in the air shower.
EUSO operates as a space-based eye looking for tracks
in one gigantic cloud chamber: the Earth's atmosphere.
The lens' opening angle  will be $60^\circ$, giving the
instrument an enormous field of view (FOV),
$\pi\times (\tan 30^\circ\times 400\ {\rm km})^2 =1.7 \times 10^5\ {\rm km}^2$.
Since the atmospheric density decreases exponentially with altitude with an 8 km
scale-height, the EUSO FOV encompasses an equivalent of $1.4\times 10^6 \ {\rm
km}^3$ of air at surface density, corresponding to
more than a Teraton of mass.

Due to the thin height of the Earth's atmosphere,
the distance $d$ between the EUSO lens and the observable air shower is highly
constrained.  For a shower directly below the lens,
$d = 400 \ {\rm km}$ to within a few percent;  for an event on the periphery of
the FOV, $d$ increases by only $1/\cos 30^\circ = 1.15$.
Thus, the energy threshold for EUSO, determined by the $1/d^2$ fall-off
of signal, is quite sharp.  In its presently planned configuration,
the threshold of EUSO is a few times $10^{19}$~eV, fortuitously positioned to
observe the region at and above the GZK cut-off.

The duration of the EUSO experiment is nominally three years,
but extensions beyond that seem likely.
On the other hand, the smaller Auger Project \cite{bib:auger1} will likely 
collect data for a decade, so that its exposure (product of acceptance and time)
may come to rival that of EUSO.
Unfortunately, the present configuration for Auger is limited to the
Southern Hemisphere, casting a blind eye in the direction of Virgo.
Other existing cosmic ray facilities  are not competitive with
the size of the EUSO FOV. 
The angular resolution expected for EUSO showers is about one degree,
or 20~mrad.  Other neutrino telescopes also offer similar resolutions.
As we shall show later, the C$\nu$B ``halo'' expected for the Virgo cluster
spans several to ten degrees,
so a one-degree resolution is sufficiently fine for the task at hand.

The paper is structured as follows.  In section \ref{sec:emission} we describe
the calculational procedure for the $Z$-burst fluxes.
Section \ref{sec:nu} discusses the modelling of the C$\nu$B density 
distribution.  Our predictions for the $Z$-burst fluxes originating from
the Virgo cluster and the corresponding event rates expected at EUSO are
presented in sections \ref{sec:fluxes} and \ref{sec:events} respectively.
We conclude in section \ref{sec:conclusion}.

\section{Preliminaries \label{sec:emission}}

The differential flux in particle species $\psi$ ($\psi=\gamma,p+n$),
\begin{equation}
F_{\psi|Z} (E, \theta,\phi) \equiv \frac{d N_{\psi|Z}}{d E \ d A \ d t \ d 
\Omega},\end{equation}
denotes the number of $\psi$ particles arriving at Earth with energy $E$
in the direction $\{\theta,\phi\}$, where $\{0,0\}$
labels the centre of the Virgo cluster (i.e., galaxy M87),
per unit energy per unit area $A$ per unit time $t$ per unit solid angle 
$\Omega$.  In the $Z$-burst scenario, this is given by
\cite{Fodor:2002hy}
\begin{eqnarray}
\label{eq:flux}
F_{\psi|Z} (E,\theta,\phi)  &= &
\sum_i \int^{\infty}_0 \!\! d E_{\psi}
\int^{R_{\rm max}}_0 \!\! dr
 \int^{\infty}_0 \!\! d E_{\nu_i} \nonumber \\
&& \times \ F_{\nu_i}(E_{\nu_i},r) \ n_{\bar{\nu}_i} (r,\theta,\phi)
 \nonumber \\ &&  \times \
\sigma_{\nu_i \bar{\nu}_i} (s) \ {\rm Br}(Z \to {\rm hadrons}) \ \frac{d 
N_{\psi}}{d E_{\psi}} \nonumber \\&& \times \  \left|\frac{\partial P_{\psi} 
(r,E_{\psi};E)}{\partial E}\right| 	+ (\nu_i \leftrightarrow \bar{\nu}_i),
\end{eqnarray}
where $F_{\nu_i} (E_{\nu_i},r)$ is the $i$th mass UHEC$\nu$ flux\footnote
{Technically, one should write $F_{\nu_i}$ as $\sum_\beta |U_{\beta 
i}|^2\,F_{\nu_\beta}$, where $|U_{\beta i}|^2=|<\nu_\beta|\nu_i>|^2$ is the
projection probability for flavour eigenstate $\beta$ onto mass eigenstate $i$,
and $F_{\nu_\beta}$ is the initial flux of flavour $\beta$.
However, this exactness introduces an unnecessary layer of detail for the 
investigation performed here.} at energy $E_{\nu_i}$ at the
$Z$ production point at a ``look-back'' time $t=r/c$,
$n_{\nu_i}(r,\theta,\phi)$ the
C$\nu$B  number density, $\sigma_{\nu_i \bar{\nu}_i}(s)$ the $Z$ production 
cross section at centre-of-mass energy
$\sqrt{s} = \sqrt{2 m_{\nu_i} E_{\nu_i}}$, ${\rm Br}(Z \to {\rm hadrons})=(69.89
\pm 0.07) \ \%$ the branching ratio,
$d N_{\psi}/d E_{\psi}$ the energy distribution of the produced $\psi$ particles
with energy $E_{\psi}$, and the propagation function $P_{\psi}(r,E_{\psi};E)$
gives the expected number of $\psi$ arriving at Earth with energies above the
threshold $E$ per particle created at $r$ with energy $E_{\psi}$.\footnote{In
the process of becoming nonrelativistic, the C$\nu$B neutrinos lose their
handedness
and depolarise to populate all of their spin states.
If neutrinos are Majorana
particles, as is predicted by the most neutrino mass models, then all C$\nu$B
spin states will participate in annihilation, and equation (\ref{eq:flux}) is
correct as written.  However, if neutrinos are Dirac particles, then half of the
nonrelativistic spin states are ``sterile'' states, and the rate presented in
(\ref{eq:flux}) should be reduced by a half.}

Since the cross section $\sigma_{\nu_i \bar{\nu}_i}(s)$ is sharply peaked at the
resonance energy $s=m_Z^2$, we may approximate
\begin{equation}
\int^{\infty}_0 d E_{\nu_i} F_{\nu_i} (E_{\nu_i}) \sigma_{\nu_i \bar{\nu}_i} (s) 
\simeq E_{\nu_i}^{\rm res}F_{\nu_i} (E_{\nu_i}^{\rm res}) \langle \sigma_{\rm 
ann} \rangle,
\end{equation}
where the superscript ``res'' denotes resonance, and
$\langle \sigma_{\rm ann} \rangle = \int ds \ \sigma_{\rm ann}/m_{Z}^2 = 40.4 \
{\rm nb}$ is the energy-averaged $s$-channel $Z$-exchange annihilation cross
section.   The UHEC$\nu$ fluxes are modelled as
\begin{equation}
F_{\nu_i} (E_{\nu_i},r) = F_{\nu_i} (E_{\nu_i},0) [1+z(r)]^{\alpha},
\end{equation}
given our insufficient knowledge about their sources.  Here,
$F_{\nu_i} (E_{\nu_i},0)$ is the neutrino flux incident on Earth,
the index $\alpha$ characterises the sources' cosmological evolution, and the
redshift $z(r)$ is related to the look-back time via $dz = -(1+z) H(z) \ dr/c$,
with $H(z)=H_0 \ [\Omega_m (1+z)^3 + \Omega_{\Lambda}]^{1/2}$ linking the Hubble
parameter at $z$ to its present value, $H_0 = h \ 100 \ {\rm km} \ {\rm s}^{-1} 
\ {\rm Mpc}^{-1}$.  Furthermore, we take $n_{\nu_i} (r,\theta,\phi) \simeq
n_{\bar{\nu}_i} (r,\theta,\phi)$,\footnote{It is interesting that the presence 
of a neutrino--antineutrino asymmetry increases both  
$n_{\nu_i}+n_{\bar{\nu}_i}$ and $|n_{\nu_i}-n_{\bar{\nu}_i}|$, while at the same
time driving one of $n_{\nu_i}$ or $n_{\bar{\nu}_i}$ to zero exponentially.  This
has the curious consequence of increasing the $Z$-burst rate, but not necessarily
the $Z$-dip depth.  For the latter, the severe suppression of $n_{\nu_i}$ or
$n_{\bar{\nu}_i}$ in the C$\nu$B gives an asymptotic depth of one half for the
$Z$-dip.  Of course, the increased number of events concomitant with a 
neutrino--antineutrino asymmetry improves the statistics of the dip.}
since significant relic neutrino--antineutrino asymmetries ($ \gwig 1$) are
incompatible with big bang nucleosynthesis in the presence of bi-large mixing
inferred from the solar and atmospheric neutrino experiments 
\cite{bib:lunardini}.\footnote{Constraints on the cosmological $\nu_\mu$ and
$\nu_\tau$ neutrino--antineutrino asymmetries are applicable insofar as 
large-angle $\nu_e \leftrightarrow \nu_{\mu,\tau}$ oscillations are operational 
prior to neutrino decoupling.  An obvious way to evade these bounds is to 
suppress these oscillations by way of new, non-standard matter effects.
One such scenario is presented in \cite{Dolgov:2004jw}, in which the suppression
arises from a hypothetical flavour-dependent neutrino--majoron coupling.} Thus,
together with the definition $F_{\nu_i}^{\rm res} \equiv F_{\nu_i}
(E_{\nu_i}^{\rm res},0) + F_{\bar{\nu}_i} (E_{\bar{\nu}_i}^{\rm 
res},0)$, equation (\ref{eq:flux}) now becomes
\begin{eqnarray}
\label{eq:flux2}
F_{\psi|Z} (E,\theta,\phi) \!&=& \!  \sum_i 2 \ {\rm Br} (Z \to {\rm hadrons}) \
\langle \sigma_{\rm ann} \rangle \ F_{\nu_i}^{\rm res}
\nonumber \\
 && \times \int^{\infty}_0 \! \! d E \int^{R_{\rm max}}_0 \! \! d r \
(1+z)^{\alpha} \ n_{\nu_i} (r,\theta,\phi)    \nonumber \\
&&  \times \ {\cal Q}_{\psi}(y)\
\left|\frac{\partial P_{\psi} (r,E_{\psi};E)}{\partial E}\right|\,.
\end{eqnarray}
The functions ${\cal Q}_{\psi}(y)=E_{\nu}/2 \cdot d N_{\psi}/d E_{\psi}$, with
$y=4 m_{\nu} E_{\psi}/m_Z^2$, are the boosted momentum distributions from
hadronic $Z$-decay, normalised to $\langle N_{p+n} \rangle = 2.04$ for 
$\psi=p+n$, and to $\langle N_{\gamma} \rangle = 2 \langle
N_{\pi^0} \rangle + \langle N_{\pi^{\pm}} \rangle = 37$ for
$\psi=\gamma$.\footnote{The photon count includes also electrons and positrons
from charged pion decay.  These are relevant for the development of
electromagnetic cascades.}  Detailed forms for ${\cal Q}_{\psi}(y)$ can be found
in reference \cite{Fodor:2002hy}.

\begin{figure*}[htp]
\epsfxsize=17.5cm
\epsfbox{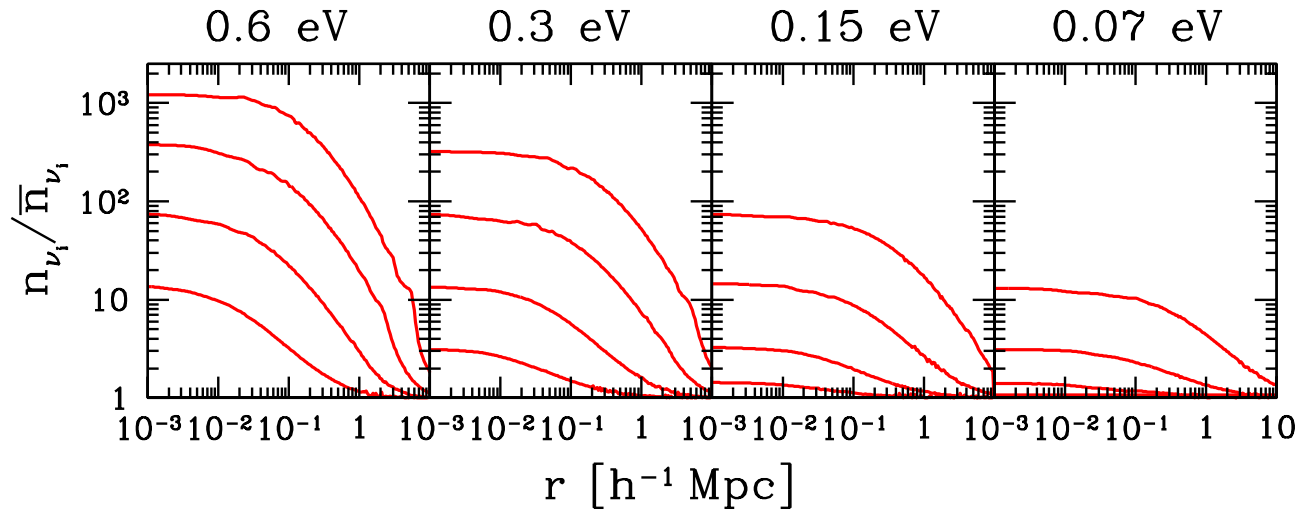}
\caption{Relic neutrino number density per mass eigenstate,
$n_{\nu_i}=n_{\bar{\nu}_i}$, as a function of the radial distance from the
halo centre, for halo virial masses (top to bottom)
$10^{15},\ 10^{14},\ 10^{13},\ 10^{12}$,
in units of $(0.7/h)\,M_{\odot}$,
and neutrino masses indicated in the figure.
All curves are normalised to the expected mean density
$\bar{n}_{\nu_i} = \bar{n}_{\bar{\nu}_i} \simeq 56 \ {\rm cm}^{-3}$.
 \label{fig:ov_dens}}
\end{figure*}

The propagation functions $P_{\psi}(r,E_{\psi};E)$ account for the interactions 
encountered by $\psi$  between its production point and Earth.  For $\psi=p+n$,
energy loss arises primarily from pion and $e^+e^-$ production
through nucleon scattering on the CMB.
The corresponding  $P_{p+n}(r,E_{p+n};E)$ has been calculated in detail in
\cite{Fodor:2000yi}, and is publicly available at \cite{bib:uhecrweb}.
For the computation of $P_{\gamma}(r,E_{\gamma};E)$,
 we adopt the continuous energy loss approximation, which asserts that the 
photon energy degradation proceeds as
\begin{equation}
dE = - E \left[ \frac{dr}{\ell_z(E)} - \frac{dz}{1+z} \right],
\end{equation}
in which $\ell_z(E) = (1+z)^{-3} \ell_0(E(1+z))$, and $\ell_0(E)$ is the photon 
energy attenuation length due to pair and double-pair production on the diffuse
extragalactic photon background, and inverse Compton scattering of the produced
pairs.  Values for $\ell_0(E)$ incorporating various assumptions about the 
poorly known URB can be found in \cite{Lee:1996fp}
and are summarised in \cite{Fodor:2002hy}. Furthermore, we assume the number of
photons $N_{\gamma}$ to be constant at energies $\gwig 10^{18} \ {\rm eV}$ due to
the small inelasticities in this energy range.
Below $\sim 10^{18} \ {\rm eV}$,
$N_{\gamma}$ increases as
$dN_{\gamma}=N_{\gamma} dr /\ell_z(E)$, so as to maintain energy conservation
(excluding losses due to the universal expansion).

\section{Neutrino density distribution \label{sec:nu}}

The C$\nu$B density distribution in the universe is taken to be uniform and at
the cosmological average $\bar{n}_{\nu_i} = \bar{n}_{\bar{\nu}_i}$ ($\simeq 56 \
{\rm cm}^{-3}$ at $z=0$), except in the vicinity of a galaxy cluster, in which
case we use
the neutrino  number densities provided in reference \cite{Ringwald:2004np}
(reproduced here in Figure \ref{fig:ov_dens}).
These densities are obtained by solving, with a particle realisation, the Vlasov 
and Poisson equations,
\begin{eqnarray}
\frac{\partial f}{\partial \tau} + \frac{\bm p}{a m_{\nu_i}} \cdot 
\frac{\partial f}{\partial {\bm x}}- a m_{\nu_i} \nabla \phi \cdot 
\frac{\partial f}{\partial {\bm p}}=0, \\\nabla^2 \phi = 4 \pi G a^2 
[\rho_m({\bm x},\tau) - \bar{\rho}_m(\tau)]\,.
\end{eqnarray}
Here,
$f(\bm{x},\bm{p},\tau)$ is the neutrino phase space distribution,
where ${\bm x}$, ${\bm p}$, and $\tau$ are the usual comoving coordinates,
conjugate momentum, and conformal time, respectively;
$a$ is the scale factor, $\bar{\rho}_m$ the mean universal matter density, and 
we assume a $\Lambda$CDM cosmology with
$\{\Omega_m,\Omega_\Lambda,h\}=\{0.3,0.7,0.7\}$.  The halo density profile is 
taken to be of the Navarro--Frenk--White form
\cite{Navarro:1995iw},
\begin{equation}\rho_m(r) =
\frac{\rho_s}{(r/r_s)(1+r/r_s)^2}\,,
\end{equation}
where $r$ is the radial distance from the halo centre, and the parameters
$r_s$ and $\rho_s$ are determined by the halo's virial mass $M_{\rm vir}$ and 
concentration $c$ via
\begin{eqnarray}
\rho_s &=& \frac{200}{3} \frac{c^3}{\ln (1+c) - c/(1+c)}, \\
r_s &=& \frac{1}{c} \left(\frac{3}{800 \pi} \frac{M_{\rm vir}}{\bar{\rho}_m} 
\right)^{1/3}.
\end{eqnarray}
In addition, high resolution simulations of reference \cite{bib:bullock2001}
provide a tight correlation between $c$ and $M_{\rm vir}$,
\begin{equation}
\label{eq:cmvir}
c(z) \simeq \frac{9}{1+z} \left(\frac{M_{\rm vir}}{1.5 \times 10^{13} h^{-1} 
M_{\odot}}\right)^{-0.13}\,.
\end{equation}
The net result is a dependence of the halo density profile on the halo's virial
mass alone.

\begin{figure*}[htp]
\epsfxsize=17.5cm
\epsfbox{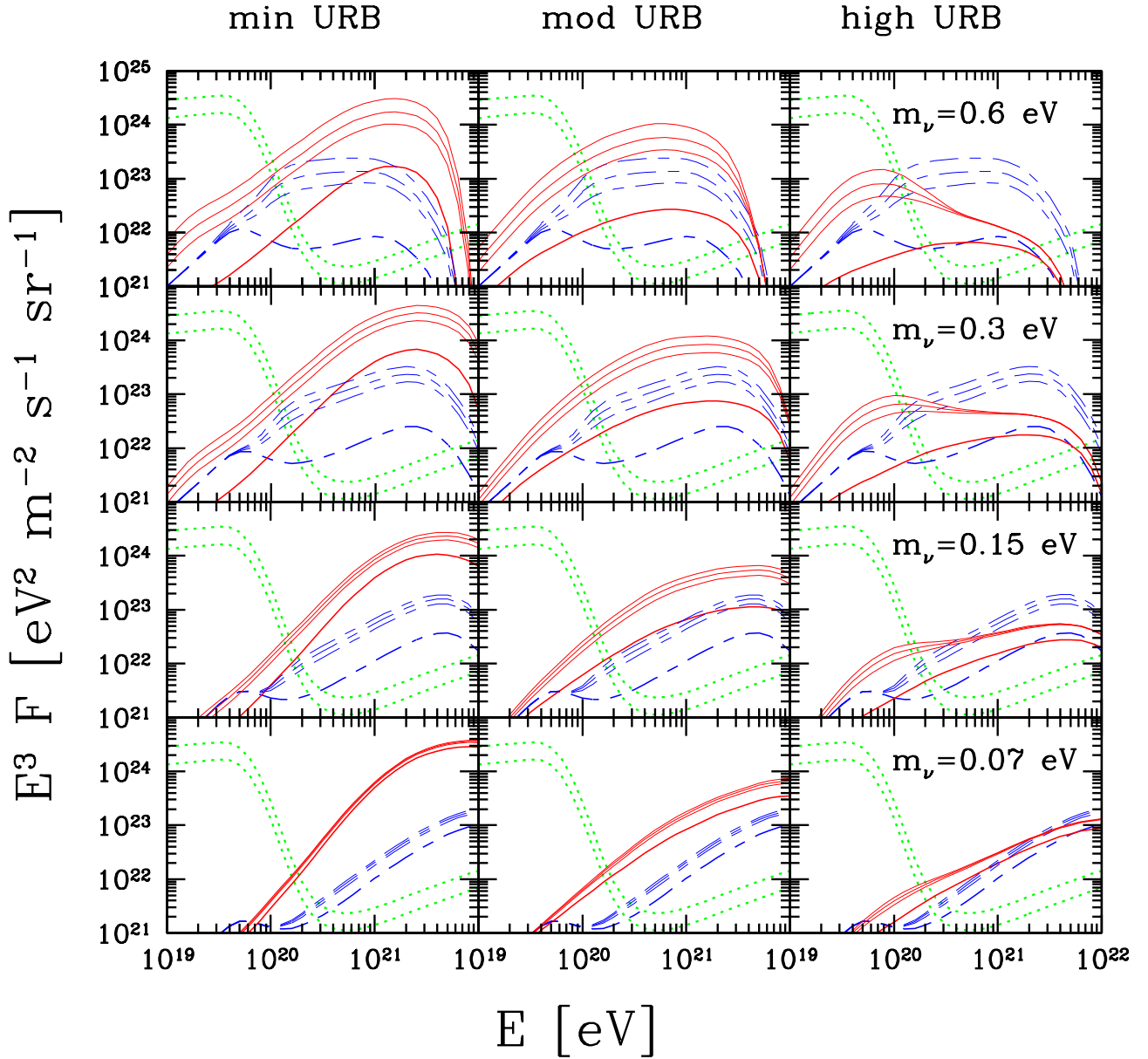}
\caption{$Z$-burst nucleon and photon fluxes from the Virgo cluster ($M_{\rm
vir} = 10^{15} M_{\odot},\ D = 15 \ {\rm Mpc}$) per neutrino species for three
URB scenarios (minimal, moderate, and high), assuming the EM cascade limited
UHEC$\nu$ fluxes. Red (solid) lines represent photon fluxes originating from
points located at angular distances  (top to bottom) $\theta=0^{\circ},4^{\circ},
10^{\circ}$, and $180^{\circ}$ (i.e., an unclustered C$\nu$B) from galaxy M87.
Blue (dashed) lines denote the corresponding nucleon fluxes. The Akeno+AGASA 
(upper) and Fly's Eye+HiRes (lower) normalised extragalactic proton fluxes are
shown in green (dotted lines).\label{fig:virgo_fluxes}}
\end{figure*}

\section{$Z$-burst fluxes from Virgo \label{sec:fluxes}}

To estimate the $Z$-burst fluxes from the Virgo cluster, we assume 
the cluster's virial mass to be $10^{15} M_{\odot}$, centred
on galaxy M87 at a distance $D \sim 15 \ {\rm Mpc}$ from Earth. 
We integrate equation (\ref{eq:flux2}) along 
the line of sight in the direction $\{\theta,\phi\}$.
The upper integration limit is taken to be
$R_{\rm max}=3000 \ {\rm Mpc}$, although the choice of $R_{\rm max}$ has little 
impact on the results, provided
$R_{\rm max}$ exceeds the GZK distance~$\sim 50 \ {\rm Mpc}$.
We note that contributions from gravitational clustering in the Milky Way halo 
($\sim 10^{12} M_{\odot}$) are generally negligible
despite its proximity.  
This is because enhancements in the C$\nu$B density therein are no more than a 
factor of twenty even for
the most massive neutrino considered here, and regions of substantial 
overdensity ($n_{\nu_i}/\bar{n}_{\nu_i} \gwig 2$) are limited in extent ($\!
\lwig 100 \ {\rm kpc}$).

Since the C$\nu$B distribution is not uniform inside the Virgo cluster but 
decreases with the halo radius, we expect the $Z$-burst fluxes to vary with
angular distance $\theta$ (recall that $\{\theta=0,\phi=0\}$ labels the centre
of the cluster) in the same manner.  Figure \ref{fig:virgo_fluxes} shows the 
predicted nucleon and photon fluxes
as functions of $E$ and $\theta$ for a range of neutrino masses and three
different URB scenarios documented in the literature.
The choice of $\alpha$ (i.e., the source evolution parameter) is irrelevant for  
Virgo $Z$-bursts, since the Virgo--Earth distance corresponds to a mere $z\sim
0.003$; we take $\alpha=0$.

For definiteness, we have opted to evaluate the UHEC$\nu$ fluxes $F_{\nu_i}^{\rm 
res}$ at the electromagnetic (EM) cascade limit, although fluxes larger by a
factor of five to twenty-five, depending on the neutrino mass, are permitted by
observations. The EM cascade limit applies generally to transparent sources
wherein neutrinos are produced in a chain of particle decays
(e.g., pions, $W$, and $Z$).  The decay process inevitably generates also
photons and/or electrons with an energy fluence comparable to that of the
neutrinos.  Subsequent EM cascades via collisions with the diffuse extragalactic
photon background, notably the CMB, bring
the cascade photons into the energy range $30 \ {\rm MeV} \div 100  \ {\rm GeV}$ 
 probed by EGRET \cite{Sreekumar:1997un}.
Observation of the diffuse extragalactic $\gamma$-ray background at these 
energies therefore places a convenient upper bound on the diffuse UHEC$\nu$
fluxes~\cite{Berezinsky:1975}.  This is known as the EM cascade limit (or
cascade limit, for short).\footnote{The EM cascade limit presented in
 \cite{Mannheim:1998wp} (displayed here in Figure \ref{fig:road})
derives from the
estimate of the diffuse $\gamma$-ray background published in
\cite{Sreekumar:1997un}.  Recent works \cite{Strong:2003ex} suggest that the
 latter analysis  may have overestimated the extragalactic
 contribution to the
$\gamma$-ray background by roughly a factor of two.  The EM cascade limit may
therefore be stronger correspondingly.}

For comparison purposes, we include in Figure \ref{fig:virgo_fluxes} also the 
spectrum of cosmic ray protons
produced at diffuse extragalactic sources,
\begin{eqnarray}
F_{p|{\rm bkd}}(E;A,\beta) \!\! &=&  \!\!  \int^{\infty}_0 \! d E_p \int^{R_{\rm 
max}}_{R_{\rm min}} \!\! d r (1+z)^{n} \ A  \left(\frac{E_p}{\rm eV} 
\right)^{-\gamma}\nonumber \\
 && \times \ \left|\frac{\partial P_p(r,E_p;E)}{\partial E}\right|\,,
\end{eqnarray}
where $\{A,n,\gamma\}=\{2.37 \times 10^{-10} \ {\rm eV}^{-1} {\rm m}^{-3} {\rm 
s}^{-1},3.65,2.54\}$ and $\{1.25 \times 10^{-10} \ {\rm eV}^{-1} {\rm m}^{-3} 
{\rm s}^{-1},3.45,2.54\}$ are the best fit parameters for  the
existing Akeno+AGASA and Fly's Eye+HiRes
data, respectively, in the energy range $10^{17.6} \div 10^{20} \ {\rm eV}$ 
\cite{Ahlers:2005sn}.  A lower integration limit of $R_{\rm min}=50 \ {\rm Mpc}$
has been imposed in the evaluation of $F_{p|{\rm bkd}}$, since
no known sources reside within this distance.  
The resulting spectrum exhibits an accumulation at the GZK scale
$4 \times 10^{19} \ {\rm eV}$, and a sharp drop beyond.

The GZK~suppression of nucleons from distant sources provides a clean 
environment for the study of $Z$-burst events originated by cosmic neutrino
messengers.  For the cascade limited UHEC$\nu$ fluxes assumed here, it is clear
from Figure \ref{fig:virgo_fluxes} that $Z$-burst events dominate above
$10^{20}$~eV, especially in the direction of Virgo.  Note that substantial
enhancements in the $Z$-burst fluxes due to C$\nu$B clustering on Virgo can still
be seen at $\theta \sim 10^{\circ}$, well beyond the cluster's visible region
$\theta \lwig 5^{\circ}$.  This follows from the assumption of an
extended CDM halo, which accumulates neutrinos in the outer region 
gravitationally despite its seeming invisibility.

\begin{figure}[t]
\epsfxsize=8.5cm
\epsfbox{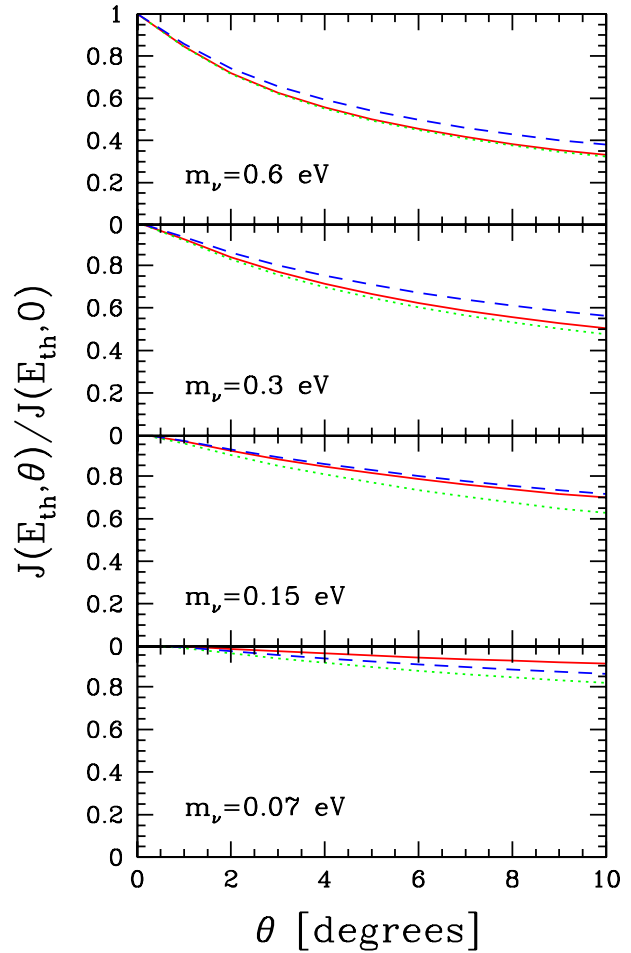}
\caption{Energy-integrated $Z$-burst fluxes (nucleons plus photons) above
$E_{\rm th}=2 \times 10^{20}\ {\rm eV}$ as functions of angular distance
from the centre of Virgo, normalised to their corresponding values at
$\theta=0^{\circ}$.  Red (solid), blue (dashed) and green (dotted) lines
denote, respectively, the minimal, the moderate and the high URB
scenarios.\label{fig:intflux}}
\end{figure}

In Figure \ref{fig:intflux} we present the ``sky map'' of the energy-integrated 
$Z$-burst fluxes for nucleons plus photons above $E_{\rm th}= 2\times
10^{20}$~eV, centred on the Virgo cluster:
\begin{equation}
J(E_{\rm th},\theta) = \sum_{\psi=p+n,\gamma} \int_{E_{\rm th}}^{\infty} d E \ 
F_{\psi|Z}(E,\theta).
\end{equation}
The shape of this curve is not very sensitive to the choice of the URB, but is 
highly dependent on the neutrino mass, since the latter determines the amount of
gravitational clustering available to the C$\nu$B.  In principle, the angular
distribution and the energy dependence of the $Z$-burst events provide
independent confirmation of the neutrino mass (and a consistency check of the
$Z$-burst mechanism).  The angular resolution of EUSO and other UHECR
detectors is typically one degree, sufficient to map out the shape of the
 extended Virgo halo.

\section{Expected events at EUSO \label{sec:events}}

The EUSO experiment detects UHECR by observing near-UV
fluorescence emitted by nitrogen molecules in
the extensive air showers generated by the
primary cosmic ray particles' interactions with the Earth's atmosphere.
For primary nucleons and photons, the cross section for interaction
is sufficiently large that the interaction
takes place high in the atmosphere with near unit probability.
Thus, the acceptance of EUSO for nucleons and photons,  ${\cal A}$,
 is just the projected FOV normal to the direction of the source, i.e.,
$\frac{1}{2} \times$ FOV $\sim 0.85\times 10^5\ {\rm km}^2$,
times the solid angle on the sky of the emitting region,
times another factor $\frac{1}{2}$ accounting for the blockage of the
upcoming beam by the opaque Earth,
times the duty cycle of the
instrument, i.e., the fraction of time ``on''.
The EUSO instrument can record fluorescence signals only on moonless
nights devoid of high cirrus clouds.  The duty cycle for such
clarity is estimated to be 20  \%.
This amounts to an all-sky acceptance of
${\cal A}_{4 \pi}= 1.1 \times 10^5\ {\rm km}^2 \ {\rm sr}$, including the duty 
cycle factor.  The nominal duration of the EUSO experiment ${\cal T}$ is three
years ($\sim 10^8 \ {\rm s}$),  so that an all-sky exposure of
${\cal E}_{4 \pi} = {\cal A}_{4 \pi} {\cal T} \simeq 1.1 \times 10^{13}
\ {\rm km}^2 \ {\rm s} \ {\rm sr}$
is anticipated.
If the lifetime of EUSO is extended, the exposure will be proportionately 
larger.

\begin{figure*}
\epsfxsize=17.5cm
\epsfbox{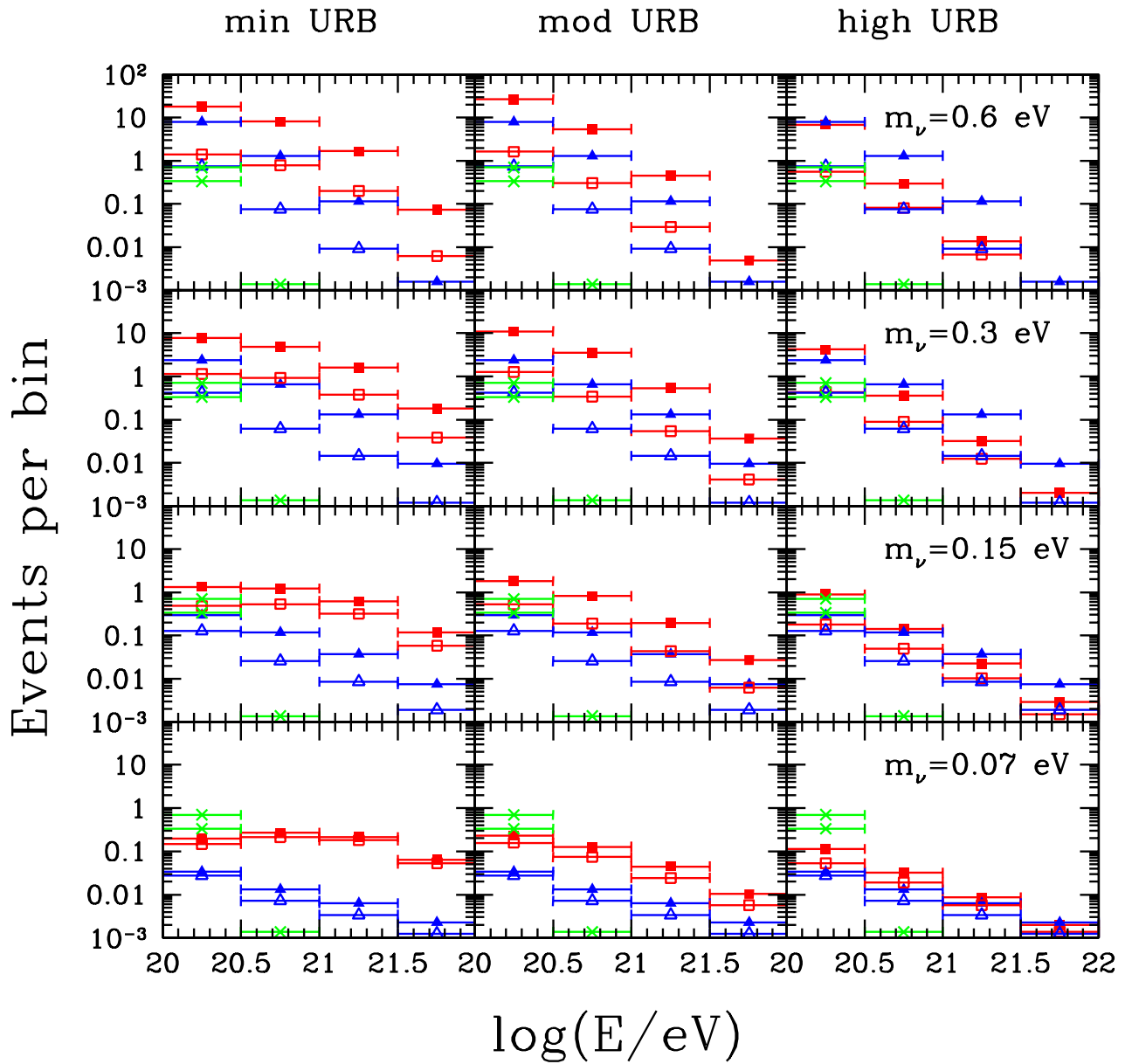}
\caption{Number of $Z$-burst events per neutrino species originating within
$\theta=10^{\circ}$ from M87 expected at EUSO in three years, assuming UHEC$\nu$ 
fluxes at the current observational limits. Solid red squares denote the photon
events, while solid blue triangles refer to the nucleons.  The corresponding 
predictions for an unclustered C$\nu$B in the same solid angle $\Delta \Omega 
\simeq 0.1 \ {\rm ster}$ are indicated by the open red squares and open blue
triangles.  The number of extragalactic background protons anticipated
 in $\Delta \Omega$, assuming  the Akeno+AGASA (upper) and the Fly's Eye+HiRes 
(lower) normalisations, are represented by the green
 crosses.
   \label{fig:obs}}
\end{figure*}

\begin{figure*}
\epsfxsize=17.5cm
\epsfbox{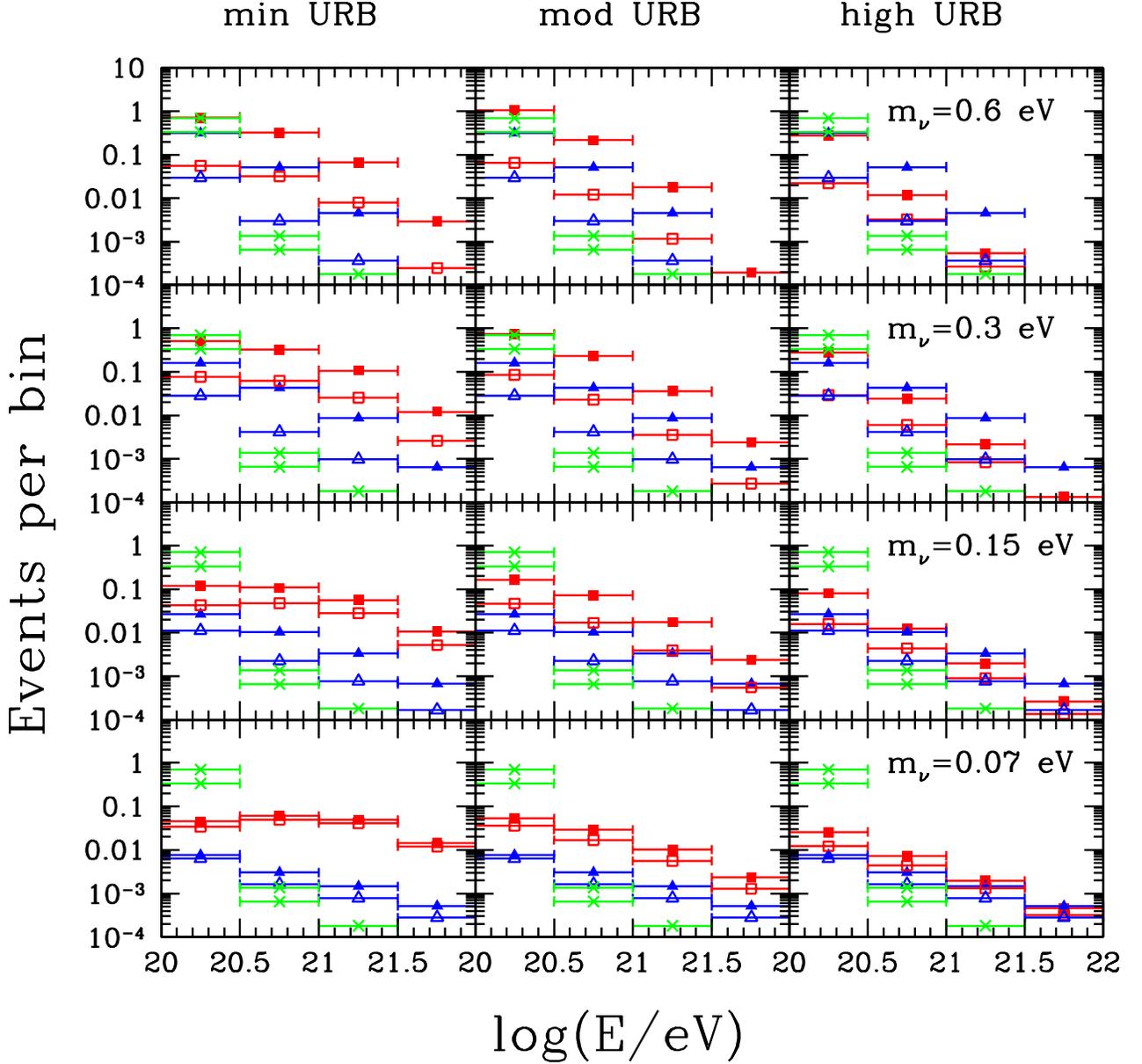}
\caption{Same as Figure \ref{fig:obs}, but for the EM cascade limited UHEC$\nu$
fluxes. \label{fig:cascade}}
\end{figure*}

For the Virgo cluster, the expected number of $Z$-burst primaries in the energy 
interval $(E_j, E_{j+1})$ originating within an angular distance $\theta$ from
the cluster centre is given by
\begin{equation}
\label{eq:npsi}
N_{\psi,j} (\theta) =  \frac{{\cal E}_{4 \pi}}{4 \pi} \! \int^{E_{j+1}}_{E_j}
\!\!\! d E \! \int^{2 \pi}_0 \!\! d \phi'
\!\! \int^{\theta}_0 \!\!
 d \theta'  \sin \theta'
F_{\psi|Z}(E,\theta',\phi').
\end{equation}
Figures \ref{fig:obs} and \ref{fig:cascade} show the integral (\ref{eq:npsi})
evaluated for $\theta = 10^{\circ}$, assuming, respectively, UHEC$\nu$ fluxes at
the current observational and the EM cascade limits,  with
$j=1, \ldots, 4$ designating the four logarithmic bins between  $10^{20} \div 
10^{22} \ {\rm eV}$. Also displayed in the figures are the numbers of 
extragalactic background protons anticipated in the same solid angle
$\Delta \Omega = 2 \pi (1-\cos \theta) \simeq 0.1 \ {\rm ster}$,
and of $Z$-burst events
in the same $\Delta \Omega$ for an unclustered C$\nu$B.

\subsection{Most optimistic scenario: hidden sources}

As is evident in Figure \ref{fig:obs}, the most favourable circumstances under 
which clustering-enhanced $Z$-burst emissions from the Virgo cluster may be
observed  occur when (i) the UHEC$\nu$ fluxes saturate the current observational
bounds, and (ii) the neutrino masses are ``large'' and quasi-degenerate,
$m_{\nu_i} \gwig 0.1 \ {\rm eV}$.\footnote
{The mass splittings inferred from the atmospheric and solar
neutrino oscillation experiments \cite{bib:atmospheric}, $\Delta m^2_{\rm atm}
\sim 2\times 10^{-3} \ {\rm eV}$ and $\Delta m^2_{\rm sun} \sim 7\times 10^{-5} \
{\rm eV}^2$, respectively, imply that the three mass eigenstates are
quasi-degenerate when $m_{\nu} \gg \sqrt{\Delta m^2_{\rm atm}}$.}
The resulting number of events at EUSO in the $\sim 0.1 \ {\rm ster}$ solid
angle can be quite large, totalling $\sim 100$ events in the energy decade
$10^{20} \div 10^{21} \ {\rm eV}$ for $m_{\nu_i} = 0.6 \ {\rm eV}$, $\sim 40$ 
for $m_{\nu_i} = 0.3 \ {\rm eV}$, and $\sim 10$ for $m_{\nu_i} = 0.15 \ {\rm
eV}$, for three degenerate species. The corresponding numbers for an unclustered
C$\nu$B are $\sim 6$, $\sim 4$, and $\sim 2$.

Theoretical models capable of generating neutrino fluxes
at these formidable energies and magnitudes---either by way of ``bottom-up''
astrophysical accelerators, or through ``top-down'' decays of super heavy
particles from beyond the standard model---are not lacking (e.g.,
\cite{Berezinsky:2003iv}). The significant hurdle facing these UHEC$\nu$ sources
is that they must be opaque to nucleons and high energy ($\gwig 100 \ {\rm MeV}$)
photons, in order not to exceed the diffuse $\gamma$-ray background observed by
the EGRET experiment
(cf.\ the EM cascade limit described in section \ref{sec:fluxes}).
A proof-of-principle candidate is
a mirror (hidden) topological defect coupled to the standard model (SM)  through
 mirror--SM neutrino oscillations; by construction, the resulting SM neutrino
fluxes is free to saturate the upper observational limits
\cite{Berezinsky:1999az}.  Furthermore, because photons are regenerated by
$Z$-decays, their subsequent EM cascade down to the energy range probed by EGRET can
lead to additional constraints on the UHEC$\nu$ sources,
particularly on the sources' cosmological evolution parameter $\alpha$.
For UHEC$\nu$ fluxes
saturating current observational limits, sources with $\alpha < 0$
are consistent with the EGRET bound \cite{bib:kalashev}.\footnote{
The EGRET constraint on $\alpha$ in \cite{bib:kalashev} applies only
to a homogeneous C$\nu$B.  However, we do not expect it to be too
seriously affected
by enhanced $Z$-burst emissions from gravitational clustering at Virgo.
This is
because the enhancements are no more than a factor of twenty, and are
concentrated in a small solid angle, $\Delta \Omega \sim 0.1 \ {\rm ster}$,
less than one hundredth of the whole sky.
Changes in the all-sky emission rate
due to enhanced emissions from Virgo can only be at the $\lwig 15 \ \%$ level.
Other galaxy clusters of comparable sizes are likely too far (i.e.,
$\gwig 50 \ {\rm  Mpc}$, outside the GZK zone) to be of great concern.}
The next generation of UHEC$\nu$ observatories will
put these neutrino source scenarios to a definitive test (cf.\ Figure \ref{fig:road}).

If neutrino masses are not quasi-degenerate, but rather hierarchical with
$m_{\nu_i} \lwig 0.1 \ {\rm eV}$,
then their clustering in Virgo is small or negligible.
For $m_{\nu_i} = 0.07 \ {\rm eV}$, the clustered and the unclustered rates 
differ generally  by no more than a factor of two
(cf.\ Figures \ref{fig:intflux} and \ref{fig:obs}).
Judging from our numbers in Figure \ref{fig:obs} (bottom row), even for an 
optimised UHEC$\nu$ flux and the best URB scenario, an experiment with at least
ten times the exposure of EUSO (three years) is required to record one 
clustering-enhanced $Z$-burst event from Virgo, another two times to record an
event within $\Delta\Omega$ with no C$\nu$B clustering,
and another five times to resolve the difference between the two at $1 \sigma$.

Of course, if there is indeed a large UHEC$\nu$ flux, it can also be measured 
directly by EUSO. It is a simple matter to obtain a good estimate of the EUSO
acceptance and exposure for neutrinos from those for cosmic ray nucleons and
photons.
To a very good approximation, the atmospheric density decreases exponentially 
vertically, with a scale-height of $x_h\sim 8$~km.  Thus, the effective volume
is $x_h \times$~FOV.  The nucleon density in this volume is
$\rho (x=0)\times N_A$, where $\rho (x=0)$ is the atmospheric density at sea
level in ${\rm g \ cm}^{-3}$, and $N_A$ is Avogadro's number.
Thus, relative to the aperture for cosmic rays, we have for neutrinos an extra
factor of $2\,x_h\,\rho(0)\,N_A\,\sigma_{\nu N}$,
where 
$\sigma_{\nu N}\approx 0.77\times 10^{-31}\,(E_\nu/10^{20} \ {\rm eV})^{0.36} \ 
{\rm cm}^2$ \cite{GQRS} is the neutrino--nucleon cross section, and the factor
of two arises because for each neutrino with an oblique trajectory, the extra
path length compensates the reduced projection of the FOV.
Putting in numbers (e.g., the vertical slant-depth, $x_h\,\rho(0)=1030 \ {\rm g
\ cm}^{-2}$, is a well known number), one arrives at
${\cal A}_{4\pi}^\nu = 0.9\times 10^{-4}\,(E_\nu/10^{20} \ {\rm 
eV})^{0.36}\,{\cal A}_{4\pi} \approx 10^{-4}\times {\cal A}_{4\pi}$ at
$E_\nu=10^{20}$~eV, with an increase in ${\cal A}_{4\pi}^\nu$ of 2.3 per decade
of energy beyond $10^{20}$~eV.  The estimate of this factor assumes that the
experimental efficiencies for cosmic rays and for neutrinos are the
same.\footnote{
This same factor describes the relative acceptances and exposures of any 
experiment triggering on downcoming and horizontal atmospheric events.  
Besides EUSO, another large area example is the Pierre Auger Observatory.
}  
For EUSO, the efficiencies for cosmic rays and for neutrinos are each
near unity at energies above $10^{20}$~eV.
The discriminator between cosmic ray and neutrino initiated events is the depth 
of the origin of the shower in the atmosphere.  For hadrons and photons, the
shower begins high in the atmosphere, while for neutrinos it begins much lower, 
where the air is densest.

In its all-sky search, EUSO should record  some
$\sim {\cal E}^\nu_{4\pi}\,E_{\nu} \,F_{\nu}(E_{\nu})$
UHEC$\nu$ events at energies above $E_\nu$.
In the energy interval $10^{21} \div 10^{22} \ {\rm eV}$,
this is of order 200  events per neutrino flavour  for $F_\nu$ at the 
observational limit, and ten times fewer for $F_\nu$ at the cascade limit.
This measurement will establish the absolute normalisation of the UHEC$\nu$ 
flux, thereby removing the final uncertainty in the $Z$-burst calculation.
It is therefore probable that clustering-enhanced $Z$-burst emissions and 
UHEC$\nu$ events will be simultaneously measured as soon as EUSO has completed 
its nominal three-year flight, if conditions (i) and (ii) are indeed satisfied.
In such a case, the associated $Z$-dips \cite{bib:absorption}
in the incident UHEC$\nu$ flux will also have been resolved by the next 
generation of dedicated UHEC$\nu$ detectors such as ANITA
in the same time frame.   
Remarkably, viable UHEC$\nu$ fluxes and neutrino masses in the $\sim 0.1$~eV 
ballpark are also able to produce the correct amount of $Z$-burst nucleons and
photons to explain the cosmic ray events observed by AGASA at energies
above the GZK cut-off.

It is interesting to note that if EUSO, or any successor experiment, succeeds
in measuring $Z$-bursts {\it and} the diffuse UHEC$\nu$ rate at $E^{\rm 
res}_{\nu_i}$, then the neutrino--nucleon cross section $\sigma_{\nu N}$ can be
inferred at $E^{\rm res}_{\nu_i} \sim 10^{22}$~eV,
far above energies available to terrestrial accelerators.
The ratio of the $Z$-burst rate and the diffuse UHEC$\nu$ rate is independent of 
the flux, but dependent on the two cross sections, $\langle\sigma_{\rm
ann}\rangle$ and $\sigma_{\nu N}$.  The former is determined purely by weak
interaction physics to be 40.4~nb,
leaving only $\sigma_{\nu N}$, which depends on the QCD dynamics of the nucleon,
as the unknown variable.

\subsection{Less optimistic scenario: transparent sources \label{sec:less}}

For transparent sources, 
the EM cascade limit on the UHEC$\nu$ fluxes applies,
thereby excluding $Z$-burst emissions as an explanation of the AGASA post-GZK 
excess (cf.\ Figure \ref{fig:road}).
Nevertheless, EUSO has a projected exposure three orders of magnitude larger 
than the exposures of existing experiments;  one may therefore hope for the
eventual discovery of $Z$-burst events in the future, even if $Z$-bursts are not
the source of the AGASA events. However, with a cascade limited UHEC$\nu$ flux,
the observation of $Z$-burst rates in the direction of Virgo seems difficult to 
realise even with three years of EUSO.  This remains true even with the
substantial clustering enhancement available to quasi-degenerate neutrinos
(Figure \ref{fig:cascade}).
Better statistics may be achievable by widening the solid angle about Virgo, 
but inevitably at the expense of narrowing the tell-tale gap between the 
clustered and unclustered rates.  Furthermore, to pursue this low signal-to-noise
scenario, one would likely require a more accurate modelling of gravitational
neutrino clustering in the local universe---something beyond our simple 
halo description.  All in all, a larger experiment is required.
The OWL and multi-OWL proposals \cite{bib:owl} would put (multi) satellites into 
orbit, to provide a FOV that dwarfs even EUSO.
Experiments using LOw-Frequency
radio antenna ARrays (LOFARs) have also been proposed.  The prototype for this 
kind of detector has just reported positive identification of UHECR \cite{LOPES}.
The Westerbork \cite{Westerbork} LOFAR facility in the Netherlands may offer the 
potential to improve the event rates of UHECR and UHEC$\nu$ by two or more orders
of magnitude beyond EUSO.

\begin{figure}[t]
\epsfxsize=8.5cm
\epsfbox{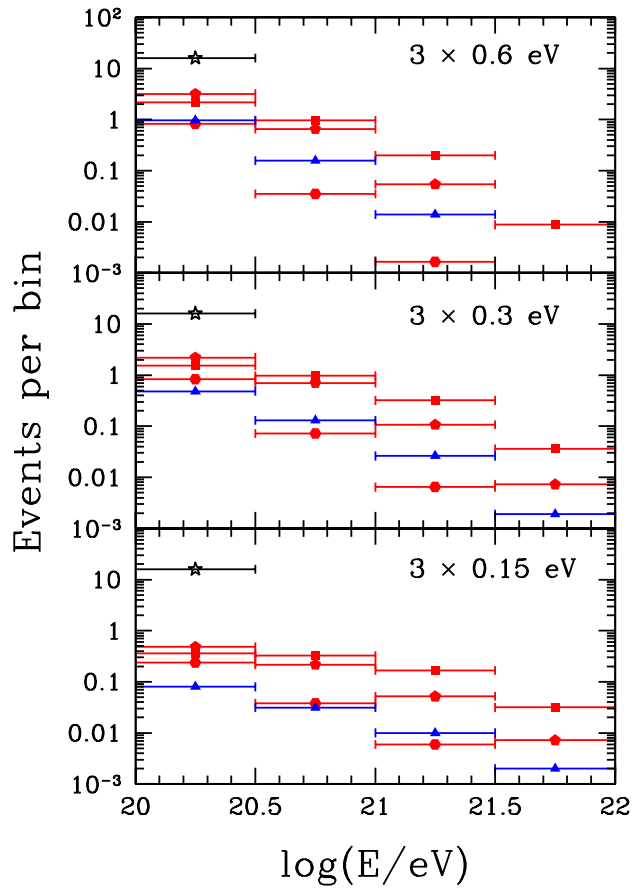}
\caption{Number of $Z$-burst events, for three quasi-degenerate neutrinos,
originating within $\theta=10^{\circ}$ from M87
 expected at EUSO in three years, assuming the cascade limited UHEC$\nu$ fluxes.
Triangles denote the nucleon events, while the squares, pentagons and hexagons
refer respectively to the photon events for the minimal, moderate and high URB
scenarios.  The stars indicate the number of events observed by AGASA in the same
energy range ($10^{20} \div 10^{20.5} \ {\rm eV}$) projected for three years of
EUSO observations in the same solid
angle.\label{fig:zburst_vs_agasa}}
\end{figure}

Suppose the UHEC$\nu$ flux comes from transparent sources, in which case it is 
cascade limited and cannot explain the post-GZK events in AGASA.  Further suppose
that the post-GZK AGASA spectrum and rate are correct.
Then, is the ``background-free'' window for neutrino physics above $E_{\rm GZK}$ 
 closed?  The answer is, not necessarily, if the C$\nu$B does cluster
sufficiently.  In Figure \ref{fig:zburst_vs_agasa}, we show
events in the energy interval $10^{20} \div 10^{21} \ {\rm eV}$ and
in a $\sim 0.1 \ {\rm ster}$ solid angle, projected
for three years of EUSO observations, 
assuming the apparent post-GZK AGASA flux (assumed to be isotropic). 
This is to be compared with the clustering-enhanced
$Z$-burst emissions from the Virgo cluster, also shown.
The latter constitute up to $3 \ \%$ and $20 \ \%$ of the
projected primary events for $m_{\nu_i} = 1.5 \ {\rm eV}$ and $0.6 \ {\rm eV}$, 
respectively, and dominate, albeit at a lower rate, beyond $10^{20.5}$~eV.
Thus, detecting $Z$-burst emissions with an observatory larger than EUSO,
even in the face of less than optimal UHEC$\nu$ fluxes and a 
possible post-GZK cosmic ray background, may not be
entirely hopeless if the C$\nu$B clusters appreciably.

\section{Conclusion \label{sec:conclusion}}

Resonant annihilation of ultra-high energy cosmic neutrinos (UHEC$\nu$) on the 
cosmic neutrino background (C$\nu$B) into $Z$ bosons---the $Z$-burst
mechanism---is a unique, albeit indirect, process capable of revealing the 
C$\nu$B in its present state.  The annihilation can be detected as absorption
dips in the incident UHEC$\nu$ flux and/or as emission features in the $Z$-decay
products (nucleons and photons) at energies above the Greisen--Zatsepin--Kuzmin
(GZK)  cut-off.  General $Z$-burst absorption and emission 
phenomenology, including the possibility that $Z$-decay products may constitute
the post-GZK ultra-high energy cosmic ray (UHECR) events observed by AGASA, has
been investigated in a number of recent publications. In the present work, we
have considered exclusively  the implications of a non-uniform C$\nu$B for the
$Z$-burst emission rates and sky map.  In particular, we have focussed on the effects of 
augmentations to the C$\nu$B number density in and around large galaxy
clusters due to gravitational clustering, and the potential of future cosmic ray
experiments, especially the Extreme Universe Space Observatory (EUSO), to observe
clustering-enhanced $Z$-burst emission rates originating from the nearby Virgo
cluster.

The GZK suppression of nucleons from extragalactic sources provides a
clean environment for the study of $Z$-burst initiated events at energies above
$\sim 10^{20} \ {\rm eV}$.  Indeed, for UHEC$\nu$ fluxes saturating the
electromagnetic cascade limit, $Z$-burst nucleon and photon fluxes originating 
from a uniform C$\nu$B already dominate at these energies. Gravitational C$\nu$B
clustering at the Virgo cluster further enhances the  fluxes by up to
a factor of several to more than forty, depending on the neutrino mass, at the 
centre of the cluster (Figure \ref{fig:virgo_fluxes}).
The enhancement decreases with angular distance from the centre in a manner 
highly sensitive to the neutrino mass (Figure \ref{fig:intflux}).  This angular
dependence of the emission events can in principle serve as an independent probe
of the neutrino mass, the shape of the Virgo halo, as well as the $Z$-burst
mechanism as a whole.

However, the statistics of $Z$-burst observation is necessarily limited by the
magnitude of the available UHEC$\nu$ flux.  For three years of EUSO observations,
we find that a detection of clustering-enhanced $Z$-burst rates from the Virgo
cluster is probable, provided that the UHEC$\nu$ fluxes are close to current 
observational limits (Figure \ref{fig:obs}).  Sources capable of
generating such large fluxes will most likely involve physics beyond the
standard model.  A quasi-degenerate neutrino mass spectrum
 is also desirable, since gravitational clustering is considerably more
efficient for ``large'' neutrino masses, $m_{\nu_i} \gwig 0.1 \ {\rm eV}$
(Figure \ref{fig:ov_dens}).  Under these favourable conditions, one would also 
expect the associated absorption dips in the incident UHEC$\nu$ flux to  be
resolved by EUSO and/or other forthcoming dedicated UHEC$\nu$ detectors.

In the case of less than optimal UHEC$\nu$ fluxes, e.g., fluxes below the
electromagnetic cascade limit, or a hierarchical neutrino mass spectrum,
the observation of enhanced $Z$-burst rates in the direction of
Virgo seems to require an experimental sensitivity exceeding that of EUSO by at
least two orders of magnitude (Figure \ref{fig:cascade}). The Westerbork radio 
observatory in the Netherlands offers a tremendous new reach in UHEC$\nu$ and
UHECR detection by looking for signals of their interactions with the lunar
regolith.  If it establishes a UHEC$\nu$ flux at energies above $10^{21} \ {\rm 
eV}$, then the final uncertainty in the $Z$-burst mechanism is removed.  This
allows definite predictions to be made for associated $Z$-burst emissions from
the Virgo cluster. Such measurements would finally establish the existence (or
not) of the cosmic neutrino background.

\acknowledgments{We thank Markus Ahlers and Birgit Eberle for useful 
discussions.
T.J.W.\ acknowledges the support of the U.S. Department of Energy
grant  DE-FG05-85ER40226 and the NASA-ATP grant 02-0000-0151.}

\end{document}